\documentstyle[12pt]{article}
\begin{document}
\vspace*{5cm}

\begin{center}
{\large UNIFICATION  OF ELECTROMAGNETISM AND GRAVITATION IN THE
FRAMEWORK OF GENERAL GEOMETRY}
\end{center}

\vspace*{.5cm}

\begin{center}
{\bf{Shervgi S. Shahverdiyev$^*$}}
\end{center}
\vspace{1.5cm}
\begin{center}
{\small   Institute of Physics, Azerbaijan National Academy of Sciences, Baku, Azerbaijan}
 \end{center}
\vspace{.5cm}

\begin{center}
{\bf{Abstract}}
\end{center}

\begin{quote}A new geometry, called General geometry, is constructed.
It is proven that its the most simplest special case is geometry underlying Electromagnetism. Another special
case is Riemannian geometry. Action for electromagnetic field and
Maxwell equations are derived from curvature function of geometry
underlying Electromagnetism. It is shown that equation of motion
for a particle interacting with electromagnetic field coincides
exactly with equation for geodesics of geometry underlying
Electromagnetism. It is also shown that Electromagnetism can not
be geometrized in the framework of Riemannian geometry.  Using
General Geometry we propose a unified model of electromagnetism
and gravitation which reproduces Electromagnetism and Gravitation
and predicts that electromagnetic field is a source for
gravitational field. This theory is formulated in four dimensional
spacetime and does not contain additional fields.
\end{quote}

\noindent
\vspace{1.5cm}

\hspace{0cm}${}^*$http://www.geocities.com/shervgis

e-mail:shervgis@yahoo.com

\newpage

\section{Introduction}
As we know equation for geodesics of  Riemannian geometry
$$
\frac{d^2x^\lambda}{du^2}=-\Gamma^{\prime\sigma}_{\lambda\nu}(x)x_u^\nu
x^\lambda,
$$
$$
2\Gamma^\prime_{\lambda,\mu\nu}=\frac{\partial{{g_{\lambda\nu}}}}{\partial
x^\mu}+\frac{\partial{{g_{\lambda\mu}}}}{\partial x^\nu}-
\frac{\partial{{g_{\mu\nu}}}}{\partial x^\lambda}.
$$
coincides with the equation of motion for a particle interacting
with gravitational field $g_{\mu\nu}$. And equation for
gravitational field is related to curvature characteristics of
Riemannian geometry
$$
^{{\bf g}}\hspace{-.09cm}S=\int dx\sqrt{-g} R^\prime
$$
$$
R^\prime=g^{\mu\nu}R^\prime_{\mu\nu},\quad
R^\prime_{\lambda\nu}=\partial_\nu\Gamma^\mu\-
_{\lambda\mu}-\partial_\mu\Gamma^\mu\-_{\lambda\nu}
+\Gamma^\mu\-_{\rho\nu}\Gamma^\rho\-_{\lambda\mu}-
\Gamma^\mu\-_{\rho\mu}\Gamma^\rho\-_{\lambda\nu},\quad
g=detg_{\mu\nu},
$$
where
$
R^\prime_{\lambda\nu}
$
is the curvature tensor of Riemannian geometry.

This can be generalized to definition of an  underlying geometry
for any theory. In the present paper,  we understand geometrization of a theory as
follows:1. Equation of motion for particle interacting with a
given field must coincide with equation for geodesics of the
corresponding geometry. 2. Equation of motion for the given field
must be related to curvature characteristics of the corresponding
geometry.

Because, Riemannian geometry and theory of gravitation satisfy
these two requirements, Riemannian geometry is considered as an
underlying geometry for gravitation.

As it is known equation or action function for gravitation could
not be found using tools of field theory because in order to get
conserved energy momentum for gravitational field it was required
to add to the action infinite number of terms. Only geometrization
principle made it possible to find a proper action for gravitation
\cite{f}.

After this was realized \cite{e}--\cite{h} at the beginning of XX
century, many physicists and mathematicians tried to geometrize
electromagnetism and unify it with gravitation using
geometrization principle \cite{w}--\cite{sid}. All these
approaches considered this problem in the framework of Riemannian
geometry and failed to satisfy the above mentioned requirements
completely or  to reproduce Electromagnetism and Gravitation
exactly. I will mention drawbacks of two well known theories,
only. They are Weyl and Kaluza-Klein theories. Drawbacks of Weyl
theory are that some of its predictions contradict experiment
\cite{p}. Drawbacks of  Kaluza-Klein theory are that it has
charge/mass problem and additional dilaton field. As it is noted
by its originator, T. Kaluza, this theory is not applicable even
to electrons because of the charge/mass problem \cite{k}.

To solve these problems, instead of increasing  dimensionality of
spacetime or choosing different metrics in Riemannian geometry, we
construct a new geometry, called General geometry. We show that it
includes Riemannian geometry as a special case, its the most
simplest special case, which have completely different properties
than Riemannian geometry, is  geometry underlying
Electromagnetism. Using its another special case we propose a
unified model of electromagnetism and gravitation which reproduces
electromagnetism and gravitation exactly  and predicts that
electromagnetic field is a source for gravitational field. It is
formulated in four dimensional spacetime and does not contain any
additional fields. Moreover, we demonstrate that Electromagnetism
can not be geometrized in the framework of Riemannian geometry,
because electromagnetic interaction has different properties than
gravitational one.

Geometry underlying the proposed model is created by interacting
particles and sources for electromagnetic and gravitational fields
unlike geometry underlying gravitation (Riemannian geometry, which
is created by sources for gravitational field only).

\section{General Geometry}

In this section we construct a new geometry. This geometry
includes  Riemannian geometry, geometry underlying
Electromagnetism (see next section), geometry underlying a unified
model of Electromagnetism and Gravitation, and infinite number of
geometries, physical interpretation of which is  not known at the
present time, as special cases. Because of this we call it General
Geometry. Besides mathematical applications, this new geometry has
important physical applications. We demonstrate this in next
section.

Let M be a manifold with coordinates $x^\lambda, \lambda=1,...,
n$. Consider a curve on this manifold $x^\lambda(u)$. Vector field
$$V=\xi^\lambda\frac{\partial}{\partial x^\lambda}$$
has coordinates 
$\xi^\lambda$. 
In  Riemannian geometry it is accepted that
\begin{equation}\label{RG}
\frac{d\xi^\lambda}{du}=-\Gamma^{\prime\sigma}_{\lambda\nu}(x)x_u^\nu\xi^\lambda,
\end{equation}
where $\Gamma^{\prime\sigma}_{\lambda\nu}(x)$ are functions of $x$
only. Because of the form of right hand side of (\ref{RG}) there
exists parallel transport of vectors along a curve. And vice
versa, if there exist parallel transport of vectors, the right
hand side of (\ref{RG}) must be in the form of
$\Gamma^{\prime\sigma}_{\lambda\nu}(x)x_u^\nu\xi^\lambda$
\cite{p}.

To construct General Geometry we do not require existence of
parallel transport and assume that
\begin{equation}\label{g}
\frac{d\xi^\sigma}{du}=-\Gamma^\sigma_\lambda(x, x_u)\xi^\lambda.
\end{equation}
$\Gamma^\sigma_\lambda(x, x_u)$ are general functions of $x$ and
$x_u$. The next step is to consider $x$ as a function of two
parameters $u, \upsilon$ and find
\begin{tabular}{l}
$\lim \Delta\xi^\sigma/\Delta u\Delta\upsilon.$\\
\vspace*{-0.8cm}
{\tiny $\Delta u\hspace*{-0.1cm}\to\hspace*{-0.1cm}0$}   \\
\vspace*{-0.19cm}
{\tiny$\Delta \upsilon\hspace*{-0.1cm}\to\hspace*{-0.1cm} 0 $} \\
\end{tabular}
In order to do that we need \vspace*{0.5cm}
$$
\frac{d\xi^\sigma}{du}=-\Gamma^\sigma_\lambda\xi^\lambda, \quad
\frac{d\xi^\sigma}{d\upsilon}=-\tilde{\Gamma}^\sigma_\lambda\xi^\lambda,
$$
$$
\Gamma^\sigma_\lambda=\Gamma^\sigma_\lambda(x, x_u,
x_\upsilon),\quad
\tilde{\Gamma}^\sigma_\lambda=\tilde{\Gamma}^\sigma_\lambda(x,
x_u, x_\upsilon).
$$
After simply calculations we arrive at
\begin{center}
\begin{tabular}{l}
$\lim\displaystyle \frac{\Delta\xi^\sigma}{\Delta
u\Delta\upsilon}=R^\sigma_\lambda\xi^\lambda,$\vspace*{-0.1cm}
\\
\vspace*{-1cm} {\tiny$ \Delta
u\hspace*{-0.1cm}\to\hspace*{-0.1cm}0$} \vspace*{.15cm}
  \\
{\tiny$\Delta \upsilon\hspace*{-0.1cm}\to\hspace*{-0.1cm} 0$}
 \\
\end{tabular}
\end{center}
where
$$
R^\sigma_\lambda=\frac{d}{d\upsilon}\Gamma^\sigma_\lambda-\frac{d}{du}
\tilde{\Gamma}^\sigma_\lambda
+\tilde{\Gamma}^\sigma_\rho\Gamma^\rho_\lambda-\Gamma^\sigma_\rho
\tilde{\Gamma}^\rho_\lambda.
$$
We call $R^\sigma_\lambda$ curvature function.

Representing $\Gamma^\sigma_\lambda(x, x_u)$ as
$$
\Gamma^\sigma_\lambda(x,
x_u)=F^\sigma_\lambda(x)+\Gamma^\sigma_{\lambda\nu}(x)x_u^\nu+
\Gamma^\sigma_{\lambda\nu\mu}(x)
x_u^\nu x_u^\mu+...
$$
and considering each order in $x_u$ or their combinations
separately we define  a set of new geometries. Only the first
order in $x_u$ is already known Riemannian geometry. Let us show
how curvature function is related to curvature tensor in the case
of Riemannian geometry.
 Let
$$
\Gamma^\sigma_\lambda(x, x_u,
x_\upsilon)=\Gamma^\sigma_{\lambda\nu}(x)x^\nu_u, \quad
\tilde{\Gamma}^\sigma_\lambda(x, x_u,
x_\upsilon)=\Gamma^\sigma_{\lambda\nu}(x)x^\nu_\upsilon.
$$
Curvature function for this case is
$$
R^\sigma_{\lambda}=R^\sigma_{\lambda\mu\nu}(x^\nu_u
x^\mu_\upsilon-x^\nu_\upsilon x^\mu_u),
$$
where
$$
R^\sigma_{\lambda\mu\nu}=\partial_\mu\Gamma^\sigma_{\lambda\nu}-\partial_\nu
\Gamma^\sigma_{\lambda\mu}
+
\Gamma^\sigma_{\rho\mu}\Gamma^\rho_{\lambda\nu}-\Gamma^\sigma_{\rho\nu}
\Gamma^\rho_{\lambda\mu}
$$
 is the curvature tensor of Riemannian geometry.

\section{Geometry of Electromagnetism}
In the case of Electromagnetism we do know  equation for the field
and particles interacting with it, but we do not know geometry
underlying it. This  is the reversed case for gravitation. We need
to know geometry underlying electromagnetism because in that case
we can construct geometry underlying unified model of
electromagnetism and gravitation  and as in the case of
gravitation, derive equation for the unified model.

For geometry of electromagnetism, we consider the most simplest
case of General Geometry
$$
\Gamma^\sigma_\lambda(x, x_u, x_\upsilon)=F^\sigma_\lambda(x(u,
\upsilon)), \quad \tilde{\Gamma}^\sigma_\lambda(x, x_u,
x_\upsilon)=F^\sigma_\lambda(x(u, \upsilon)),
$$
when $\Gamma^\sigma_\lambda(x,x_u)$ does not depend on $x_u$ and
show that it is an underlying geometry for electromagnetism. In
order to prove, that geometry  defined by
\begin{equation}\label{GE}
\frac{d\xi^\sigma}{du}=-F^\sigma_\lambda(x)\xi^\lambda
\end{equation}
with the length of a curve
$$
ds=\sqrt{\eta_{\mu\nu}dx^\mu dx^\nu}+\frac{q}{cm}A_\mu dx^\mu
$$
is an underlying geometry for electromagnetism we must show that
equation of motion for a particle interacting with electromagnetic
field coincides with equation of geodesics in this geometry, and
Maxwell equations and  Lagrangian for electromagnetic field are
related to its curvature characteristics.

Geometry defined by (\ref{GE}) has different properties than
Riemannian geometry defined by (\ref{RG}). We do not get into
details here. We simply mention that in this geometry the notion
of parallel transport is not defined. As we show in the sequel
this makes it be underlying geometry for Electromagnetism.

To obtain  equations for geodesics we substitute $\xi^\lambda$  in
(\ref{GE}) by $x^\lambda_u$ and arrive at
$$
\frac{d^2x_\sigma}{du^2}=-F_{\sigma\lambda}(x)x^\lambda_u.
$$
This is exactly equation of motion for a charged particle moving
in electromagnetic field $A_\mu$, if we choose
 $$
F_{\sigma\lambda}=\frac{q}{cm}(\partial_\sigma
A_\lambda-\partial_\lambda A_\sigma).
$$
So, the first requirement is satisfied with this choice of
function $F_{\sigma\lambda}$. In  \cite{sss}, we have proved this
relation between $F_{\sigma\lambda}$ and $A_\mu$.

It remains to show that Maxwell equations and  Lagrangian for
electromagnetic field is related to curvature characteristics of
geometry (\ref{GE}). To this end let us find curvature function
for (\ref{GE})
$$
R^\sigma_{\lambda}=R^\sigma_{\mu\lambda}(x^\mu_\upsilon-x^\mu_u),
$$
where
$$
R^\sigma_{\mu\lambda}=\partial_\mu F^\sigma_\lambda.
$$
This tensor is an analog of curvature tensor of Riemannian
geometry. After summing by two of the three indices we obtain
$$
R_\lambda=R^\mu_{\mu\lambda}=\partial_\mu F^\mu_\lambda.
$$
Vector $R_\lambda$ is an analog of Ricci tensor. Equations
$R_\lambda=0$ coincide with the Maxwell equations. In order to
construct a Lagrangian we need a scalar function. In our case we
have two quantities $R_\lambda $ and $A^\lambda$. $A^\lambda$
originates from the length of a curve (metric) as $g_{\mu\nu}$
originates from the length of a curve in Riemannian geometry. We
can construct from $R_\lambda $ and $A^\lambda$ a Lagrangian
$$
R=A^\lambda R_\lambda =\partial_\mu(A^\lambda
F^\mu_\lambda)-\frac{1}{2}F_{\mu\lambda}F^{\mu\lambda}.
$$
This coincides with the Lagrangian of electromagnetic field up to
total derivative.

We see that as in the case of Riemannian geometry and gravitation
we can find equations and action functional for electromagnetic
field from geometric characteristics of geometry underlying
Electromagnetism. And equation for geodesics coincides with the
equation of motion for a particle interacting with electromagnetic
field.

From the geometrical point of view a charged particle interacting
with electromagnetic field can be considered as a free particle in
the spacetime with the length of a curve
$ds=\sqrt{\eta_{\mu\nu}dx^\mu dx^\nu}+\frac{q}{cm}A_\mu dx^\mu$
and  equation for geodesic
$$
\frac{d^2x_\sigma}{du^2}=\frac{q}{cm}(\partial_\lambda
A_\sigma-\partial_\sigma A_\lambda)x^\lambda_u,
$$
where $A_\mu$ is a solution to  equation $R_\lambda=0$.

This theory does not have any drawbacks like in the theories
constructed before. It reproduces electromagnetism exactly, is
free from additional fields and extra dimensions.

\section{Unification of electromagnetism and gravitation}

Now, we consider a different special case of General Geometry.
For geometry underlying our unified model we choose functions
$\Gamma$ and $\tilde{\Gamma}$ as
$$
\Gamma^\sigma\-_\lambda(x, x_u, x_\upsilon)=F^\sigma\-_\lambda(x(u,
\upsilon))+\Gamma^\sigma\-_{\lambda\nu}(x)x^\nu_u,\quad
\tilde{\Gamma}^\sigma\-_\lambda(x, x_u,
x_\upsilon)=F^\sigma\-_\lambda(x(u,\upsilon))+\Gamma^\sigma\-_{\lambda\nu}(x)x^\nu_\upsilon.
$$
And the length of a curve as
$$
ds=\sqrt{g_{\mu\nu}(x)dx^\mu dx^\nu}+\frac{q}{cm}A_\mu(x) dx^\mu.
$$
For our choice of $\Gamma^\sigma\-_\lambda$, (\ref{g}) becomes
\begin{equation}\label{*}
\frac{d\xi^\sigma}{du}=-(F^\sigma\-_\lambda(x)+
\Gamma^\sigma\-_{\lambda\nu}(x)x_u^\nu)\xi^\lambda,
\end{equation}
We substitute $\xi^\sigma$ in (\ref{*}) by $x^\sigma_u$ and obtain equation for geodesics
$$
\frac{d^2x^\sigma}{du^2}=-F^\sigma\-_\lambda(x)x^\lambda_u- \Gamma^\sigma\-_{\mu\nu}(x)x^\mu_u x^\nu_u.
$$
 It coincides exactly with  equation of motion for a particle with charge $q$
 and mass $m$ interacting with electromagnetic and gravitational fields if we choose
\begin{equation}\label{2}
F_{\mu\nu}=\frac{q}{cm}(\partial_\mu A_\nu-\partial_\nu A_\mu),
\quad
2\Gamma_{\lambda,
\mu\nu}=\frac{\partial{{g_{\lambda\nu}}}}{\partial
x^\mu}+\frac{\partial{{g_{\lambda\mu}}}}{\partial x^\nu}-
\frac{\partial{{g_{\mu\nu}}}}{\partial x^\lambda}.
\end{equation}
In this paper we assume these relations and declare $A_\mu$ as electromagnetic field
 and $g_{\mu\nu}$ as gravitational field. These relations between $F$ and $A$, and
 $\Gamma$ and  $g_{\mu\nu}$ are proven in \cite{sss}
and it is  shown that $A_\mu$ can be identified with
 electromagnetic field, $q$ with  charge, $m$ with mass  of
 a particle interacting with $A_\mu$, $ c $ with the velocity of the light,
  and $g_{\mu\nu}$ with gravitational field.

The corresponding curvature function is
\begin{equation}\label{1}
R^\sigma\-_{\lambda}=(\partial_\mu
F^{\sigma}\-_\lambda-\Gamma^\rho\-_{\lambda\mu}F^\sigma\-_\rho+
\Gamma^\sigma\-_{\rho\mu}F^\rho\-_\lambda)(x^\mu_\upsilon-x^\mu_u)+
$$
$$
\frac{1}{2}(\partial_\nu\Gamma^\sigma\-_{\lambda\mu}
-\partial_\mu\Gamma^\sigma\-_{\lambda\nu}
+\Gamma^\sigma\-_{\rho\nu}\Gamma^\rho\-_{\lambda\mu}-\Gamma^\sigma\-_{\rho\mu}\Gamma^\rho\-_{\lambda\nu})
(x^\nu_\upsilon x^\mu_u-x^\nu_u x^\mu_\upsilon).
\end{equation}
From (\ref{1}), we see that gravitational field is coupled to $F^{\sigma}\-_\lambda$ through covariant derivative
$$
\Delta_\mu F^{\sigma}\-_\lambda=\partial_\mu
F^{\sigma}\-_\lambda-\Gamma^\rho\-_{\lambda\mu}F^\sigma\-_\rho+\Gamma^\sigma\-_{\rho\mu}F^\rho\-_\lambda.
$$
We have $g_{\mu\nu}$, $ A_\mu$ and curvature function to use to find an  action for the unified model.
First, we construct a tensor from
(\ref{1})\footnote{One may construct different tensors from (\ref{1}).}
$$
 R^\sigma\-_{\lambda\mu\nu}=\frac{cm}{4q}(\Delta_\nu F^{\sigma}
\-_\lambda A_\mu-\Delta_\mu F^{\sigma}\-_\lambda A_\nu)+
\frac{1}{16\pi G}(\partial_\nu\Gamma^\sigma\-_{\lambda\mu}
-\partial_\mu\Gamma^\sigma\-_{\lambda\nu}
+\Gamma^\sigma\-_{\rho\nu}\Gamma^\rho\-_{\lambda\mu}-\Gamma^\sigma\-_{\rho\mu}\Gamma^\rho\-_{\lambda\nu}),
$$
where $G$ is gravitational constant.
Finally we have a scalar
$$
 R=g^{\lambda\nu}R^\mu\-_{\lambda\mu\nu}
$$
and  action
$$
^{{\bf eg}}\hspace{-.09cm}S=\int dx\sqrt{-g} R=\frac{c^2m^2}{4q^2}
\int dx\sqrt{-g}F^{\mu\nu}F_{\mu\nu}+\frac{1}{16\pi G}
\int dx\sqrt{-g}\hspace{.1cm} {}^{{\bf g}}\hspace{-.1cm}R,
$$
$$
{}^{\bf g}\hspace{-.1cm}R=g^{\mu\nu}\hspace{.05cm}{}^{\bf g}\hspace{-.1cm}R_{\mu\nu},\quad
{}^{\bf g}\hspace{-.1cm}R_{\lambda\nu}=
\partial_\nu\Gamma^\mu\-_{\lambda\mu}-\partial_\mu\Gamma^\mu\-_{\lambda\nu}
+\Gamma^\mu\-_{\rho\nu}\Gamma^\rho\-_{\lambda\mu}-\Gamma^\mu\-_{\rho\mu}
\Gamma^\rho\-_{\lambda\nu},\quad g=detg_{\mu\nu},
$$
where  use has been made of
$$
\Delta_\nu F^{\mu\nu}=\frac{1}{\sqrt{-g}}\partial_\nu(\sqrt{-g}F^{\mu\nu}),
\quad
\Delta_\nu g_{\mu\lambda}=0.
$$
Note, that the action is invariant under gauge transformations
of fields and general transformations of coordinates. Covariant
derivative appears naturally in this formalism. Hence, geometrization
principle leads to an action which is invariant under gauge transformations
of fields and general transformations of coordinates. We conclude that
geometrization principle is more general than gauge principle.

Equation of motion for gravitational field is
\begin{equation}\label{3}
\frac{c^2m^2}{4q^2}(\frac{1}{2}F^{\rho\sigma}F_{\rho\sigma}g^{\mu\nu}-2F^{\nu\sigma}F^\mu\-_\sigma)+\frac{1}{16\pi G}(-{}^{\bf g}\hspace{-.1cm}R^{\mu\nu}+
\frac{1}{2}\-{}^{\bf g}\hspace{-.1cm}Rg^{\mu\nu})=0.
\end{equation}
From (\ref{3}) it follows that
$^{\bf g}\hspace{-.1cm}R=0$ for $n=4$ (in the rest of the paper we restrict ourselves to four dimensional spacetime)
and (\ref{3}) becomes
\begin{equation}\label{4}
\frac{c^2m^2}{4q^2}({1\over 2} g^{\rho n}g^{\sigma m}F_{nm}F_{\rho\sigma}g_{\mu\nu}-2g^{\sigma m}F_{\nu m}F_{\mu\sigma})-\frac{1}{16\pi G}{}^{\bf g}\hspace{-.1cm}R_{\mu\nu}=0.
\end{equation}
We see that electromagnetic field is a source for  gravitational field.
In the weak gravitational and strong electromagnetic field approximation  $g^{\mu\nu}\sim \eta^{\mu\nu}=diag(1, -1, -1, -1)$ and
$$
{}^{\bf g}\hspace{-.1cm}R_{00}\sim-\partial_\mu\Gamma^\mu_{00}=-\frac{1}{2}{\bf{\Delta}} g_{00}, \quad {\bf{\Delta}}=\frac{\partial}{\partial x^i}\frac{\partial}{\partial x^i}, \quad i=1,2,3.
$$
The $00$ component of equation (\ref{4})  gives
\begin{equation}\label{5}
{\bf{\Delta}} \Phi=4\pi c^2 G(E_iE_i+H_iH_i)+O(hF),\quad g_{00}=1-2\frac{\Phi}{c^2},
\end{equation}
where $\Phi$ is the Newtonian potential,
$E_i=\partial_0 A_i-\partial_i A_0$ and
$H_i=\frac{1}{2}\epsilon_{ijk}(\partial_j A_k-\partial_k A_j)$ are  electric and
magnetic fields respectively and $\epsilon_{ijk}$ is antisymmetric tensor.
Accordingly, total energy of electromagnetic field produces gravitational field.

 Because geometrization
principle gave  true equation for gravitational field \cite{f}, we
can be sure that this equation is also true. The proposed theory
gives exactly Gravitation when electromagnetic field is equal to
zero and Electromagnetism when gravitational field is equal to
zero. It predicts that electromagnetic field is a source for
gravitational field. This theory is formulated in four dimensional
spacetime and does not contain any additional fields.

\section{Discussion}

For Riemannian geometry

$$
\frac{d\xi^\lambda}{du}=-\Gamma^{\prime\sigma}_{\lambda\nu}(x)x_u^\nu\xi^\lambda,
$$
 it is possible to make a change of coordinates so that its right
hand side will be equal to zero (parallel transport), because of
its right hand side structure. In new coordinates $x^\prime$,
equation for geodesics becomes
$$\frac{d^2x^{\prime\sigma}}{du^2} =0.$$
From physical point of view this corresponds to finding a
reference frame where trajectory of particles is strait line,
because this equation must coincide with the equation of motion.
For gravitational interaction we can find a reference frame  where
gravitational interaction is absent. Accordingly, Riemannian
geometry is suitable for gravitational interaction only. For
electromagnetic interactions it is not possible to find a
reference frame  where it is absent. Therefore, all attempts to
geometrize electromagnetism or unify it with gravitation in the
framework of Riemannian geometry must fail (see also
\cite{cfgrft}).

 On the other hand for geometry
 $$\frac{d\xi^\sigma}{du}=
 -(F^\sigma\-_\lambda(x)+\Gamma^\sigma\-_{\lambda\nu}(x)x_u^\nu)\xi^\lambda
 $$
  it
is not possible to eliminate its right hand side by changing
coordinates because of $F^\sigma\-_\lambda$ term. And this
property makes it be underlying geometry for the proposed unified
model.

In general relativity, geometry underlying gravitation and metric
are independent of properties  of interacting particles. This is a
consequence of equivalence principle. Geometry and metric depends
on the characteristics of sources for gravitational field
$g_{\mu\nu}$ only. For electromagnetic interactions there is no
equivalence principle, so geometry and metric underlying
electromagnetism and unified model of electromagnetism and
gravitation must depend on characteristics of interacting
particles, because particles of different charges move in
electromagnetic field differently.

For our model we have
$$
\frac{d\xi^\sigma}{du}=-(\frac{q}{cm}(\partial_\mu A_\nu-\partial_\nu A_\mu)+
\Gamma^\sigma\-_{\lambda\nu}(x)x_u^\nu)\xi^\lambda.
$$
Accordingly, geometry underlying unified model of electromagnetism and gravitation
depends not only on the characteristics of sources for $A_\mu$ and
$g_{\mu\nu}$ but also on the characteristics of interacting particles $q$ and $m$.
 This means that geometry and the length of a curve (metric)
$
ds=\sqrt{g_{\mu\nu}(x)dx^\mu dx^\nu}+\frac{q}{cm}A_\mu(x) dx^\mu
$
are created by interacting particles too, together with sources unlike gravitational
interaction.  This gives us a new understanding of problem of geometry and matter.

Next, we note that in General Relativity, if we consider sources
for gravitational field we add the so called source term
$g_{\mu\nu}T^{\mu\nu}$ to the action
$$
^{{\bf g}}\hspace{-.09cm}S=\int dx(\sqrt{-g} R^\prime
+g_{\mu\nu}T^{\mu\nu}).
$$
$T^{\mu\nu}$ is the energy-- momentum tensor of sources for
gravitational field. This is called the "covariance principle", i.e. an assumption. In
the other words it can be rephrased as "everything that has energy
couples to gravity". Some scientists misinterpret this principle
as a prediction of General Relativity in the form "everything that
has  energy is a source for gravitational field". It is obvious
that this is not correct, because all principles (action
principle, geometrization principle, covariance principle and
etc.)i.e. assumptions can not be at the same time
predictions. Therefore, we can not replace $T^{\mu\nu}$
with energy- momentum tensor of any field, if it is not a source
for $g_{\mu\nu}$. Inclusion of electromagnetic field
in $T^{\mu\nu}$ declares it as a source for gravitational field
which is an assumption but not a prediction.

Resuming, we can say that we have eliminated the need for extra
dimensions and additional fields for proposing unified model of
electromagnetism and gravitation by formulating a new geometry.
This approach can be useful for formulating a unified electroweak
model without Higgs fields and for unifying strong interactions
with the other ones.

\end{document}